\documentclass[reprint,groupedaddress]{revtex4-2}
\usepackage{amsmath,amssymb,mathrsfs,amsfonts}
\usepackage{txfonts}
\usepackage{upgreek}
\usepackage{bm}
\usepackage{color}
\usepackage{appendix}
\usepackage{graphicx}
\usepackage{float}
\usepackage{footmisc}
\usepackage{bm}
\usepackage{color}
\usepackage{CJKutf8}
\usepackage{epsfig}
\usepackage[dvipdfm,colorlinks,
            citecolor=blue,
            anchorcolor=red,
            menucolor=red,
            linkcolor=blue,
            filecolor=red,
            runcolor=red,
            frenchlinks=red]{hyperref}

\preprint{APS/123-QED}

\begin{document}
\begin{CJK}{UTF8}{gbsn}
  \title{Laser-field detuning assisted optimization of exciton valley dynamics in monolayer WSe$_2$: Geometric quantum speed limit}

   \author{Kang Lan$^1$}
   \author{Shijie Xie$^1$}
   \email{xsj@sdu.edu.cn}
   \author{Jiyong Fu$^2$}
  \email{yongjf@qfnu.edu.cn}
\affiliation{$^1$School of Physics, State Key Laboratory of Crystal Materials, Shandong University, Jinan 250100, China\\
$^2$ Department of Physics, Qufu Normal University, Qufu 273165, China}

\begin{abstract}
  Optimizing valley dynamics is an effective instrument towards precisely manipulating qubit in the context of two-dimensional semiconductor.
  In this work, we construct a comprehensive model, involving both intra- and intervalley channels of excitons in monolayer WSe$_2$, and simultaneously takes the light-matter interaction into account, to investigate the optimal control of valley dynamics with an initial coherent excitonic state.
  Based on the quantum speed limit (QSL) theory, we propose two optimal control schemes aiming to reduce the evolution time of valley dynamics reaching the target state, along with to boost the evolution speed over a period of time.
  Further, we emphasize that the implementation of dynamical optimization is closely related to the detuning difference---the difference of exciton-laser field detunings between the $K$ and $K'$ valleys---which is determined by the optical excitation mode and magnetically-induced valley splitting.
  In particular, we reveal that a small detuning difference drives the actual dynamical path to converge towards the geodesic length between the initial and final states, allowing the system to evolve with the least time.
  Especially, in the presence of valley coherence, the actual evolution time and the calculated QSL time almost coincide, facilitating high fidelity in information transmission based on the valley qubit.
  Remarkably, we demonstrate an intriguing enhancement in evolution speed of valley dynamics, by adopting a large detuning difference, which induces an emerging valley polarization even without initial polarization.
  Our work opens a new paradigm for optically tuning excitonic physics in valleytronic applications, and may also offer solutions to some urgent problems such as speed limit of information transmission in qubit.
\end{abstract}

\maketitle

\section{introduction}
\label{sec:intr}
Along with the discovery of the direct band gap at two inequivalent $K$ and $K'$ points of the Brillouin zone, transition metal dichalcogenides (TMDCs) MX$_2$ (M=Mo, W; X=S, Se, Te) have emerged as ideal candidates for novel microscopic devices~\cite{PhysRevB.77.235406,PhysRevLett.108.196802}.
Benefiting from space-inversion asymmetry together with strong spin-orbit interaction, there forms a \emph{spin-valley-locked} electronic structure, protected by the time-inversion symmetry, which leads to opposite spin splitting at the two distinct valleys~\cite{wangkalantar2012}.
This has motivated a host of hot research topics in valley physics, such as the control over valley polarization~\cite{PhysRevB.97.115425,aivazianmagnetic2015,zeng2012valley,PhysRevB.104.195424} and valley coherence~\cite{hao_moody2016,PhysRevB.50.10868,PhysRevB.108.035419,moody2015intrinsic,selig2016excitonic}, valley Hall effect~\cite{PhysRevLett.108.196802,science1250140}, as well as valley entanglement~\cite{PhysRevB.92.075409,PhysRevB.107.035404}.
Also, owing to the large electron and hole effective masses and reduced dielectric screening, the monolayer TMDCs exhibit strong Coulomb interactions. This enables the formation of tightly bound excitons with remarkably large binding energy up to hundreds of meV~\cite{PhysRevLett.113.026803,PhysRevLett.113.076802,PhysRevB.89.125309}, allowing for excitonic valley control even at room temperature.

The \emph{spin-valley-locked} excitonic states can be formed directly by using circularly polarized light ($\sigma_+$ and $\sigma_-$), based on the optical selective rule~\cite{PhysRevLett.108.196802,cao2012,PhysRevB.90.041414,acs.nanolett.0c01019}.
Further, the intervalley transition of excitons between $K$ and $K'$ requires a spin flip for both electrons and holes~\cite{nanolett.8b01484,acs.jpcc.2c05113}.
Hence, in close analogy with spin and its associated applications in quantum information, the valley pseudospin as an additional degree of freedom, provides a fascinating platform to encode and process binary information~\cite{Qu2019,PhysRevB.97.115425,PhysRevB.107.035404}, that underlines a prospective coherent control based on \emph{spin-valley-locked} excitonic states.
And, the coherent excitonic states can be initialized, controlled, and read out on ps time scale~\cite{PhysRevLett.117.187401,PhysRevLett.117.077402}.

A fundamental requirement for information processing devices is the ability to universally control the qubit state on a time scale shorter than its coherence time.
With existing technology applications, the question naturally arises as to how to optimally manipulate these devices.
Particularly for valley-coherent exciton states, whose relatively short coherence time ($\sim$ ps) restricts the spatial transmission of information about the logical process in the valley qubit~\cite{hao_moody2016,PhysRevLett.116.127402,PhysRevLett.117.187401,dufferwiel2018}.
Despite several efforts that have been made to boost the valley coherence time, e.g. exciton-cavity coupling~\cite{qiu2019roomtemperature}, electron doping~\cite{yangsinitsyn2015}, magnetic suppression~\cite{PhysRevB.108.035419}, and enhanced dielectric screening~\cite{gupta2023}, how to optimally govern the valley dynamics that incorporates both excitons generation, intravalley recombination, intervalley transfer and coherence loss processes, to facilitate information transmission, is still not available.

\begin{figure}
\includegraphics[width=0.95\linewidth]{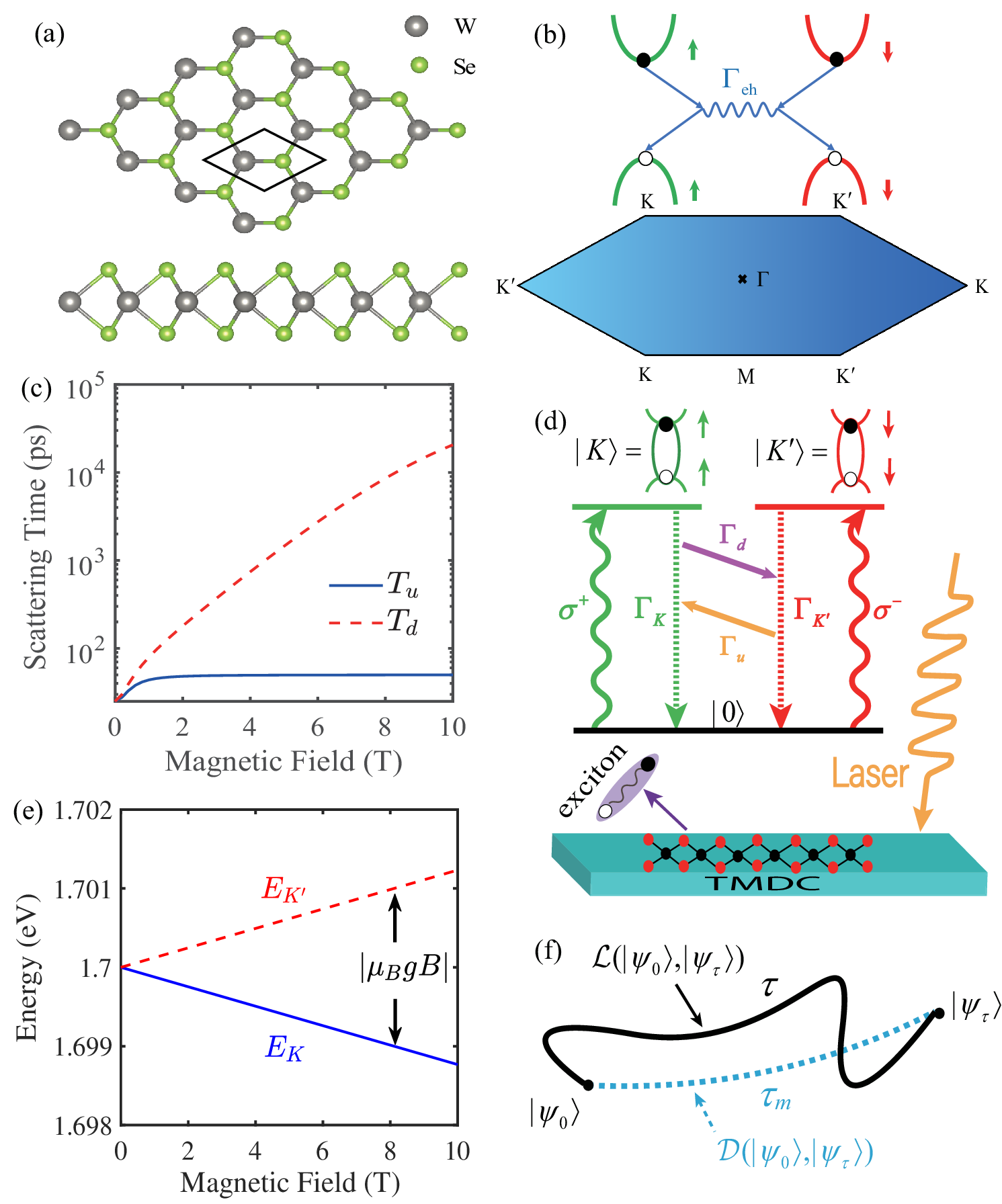}
\caption{(a) The schematic structure of top (upper pattern) and side (lower pattern) views of monolayer WSe$_2$.
(b) Upper panel: representation of incoherent intervalley transfer of excitons between $K$ and $K'$ valleys, caused by the electron-hole exchange and exciton-phonon interactions.
Black and white circles represent electrons and holes, respectively.
Lower panel: schematic diagram of Brillouin zone, where the $K$ and $K'$ valleys are associated by time-reversal symmetry.
(c) The intervalley scattering times $T_u$ and $T_d$ as a function of external magnetic field at temperature $T=4$ K.
(d) Illustration of coupled structure of the TMDC and laser field: intravalley excitation, intervalley transfer and recombination channels.
The basis set $\{|K\rangle,|K'\rangle,|0\rangle\}$ characterizing the exciton states in the $K$ ($|K\rangle$) and $K'$ ($|K'\rangle$) valleys and the ground state ($|0\rangle$).
$\sigma^{+}$ ($\sigma^{-}$) denotes the right (left) polarized light excitation, and $\Gamma_{K(K')}$ is the exciton recombination rate.
$\Gamma_{d(u)}$ is the intervalley scattering rate from $K(K')$ valley to $K'(K)$ valley.
(e) Energy levels of bright excitons in $K$ (solid line) and $K'$ (dashed line) valleys in the presence of magnetic field.
(f) Physical schematic of the geometric QSL: the length of the geodesic $\mathcal{D}(|\uppsi_0\rangle,|\uppsi_\uptau\rangle)$ defines the lower bound of actual evolution $\mathcal{L}(|\uppsi_0\rangle,|\uppsi_\uptau\rangle)$ between the initial state $|\uppsi_0\rangle$ and final state $|\uppsi_\uptau\rangle$, with $\mathcal{D}\leq\mathcal{L}$.}
\label{figure1}
\end{figure}

In this work, we construct a comprehensive model that incorporates both intra- and intervalley channels of excitons in monolayer WSe$_2$, and simultaneously consider the light-matter interaction, to endow an unambiguous scheme for optimally tuning valley dynamics.
To this end, we employ the quantum speed limit (QSL) theory, which stipulates the minimum time for a quantum system to evolve from an initial state to its distinguishable state~\cite{PhysRevLett.111.260501,Lan_2022,PhysRevLett.111.010402,PhysRevA.94.052125,PhysRevA.95.052104,PhysRevA.93.020105,PhysRevA.89.012307,PhysRevLett.103.240501}.
For linearly polarized excitation with an initial coherent superposition of excitonic states, we propose two optimal control schemes of valley dynamics with respect to different \emph{performance measures}, allowing to tune the evolution time and average speed of valley dynamics by the detuning difference.
We reveal that a small detuning difference favors the reduction of evolution time of valley dynamics, enabling the actual dynamical path to converge towards the geodesic length between the initial and final states.
In contrast, a large detuning difference leads to an enhanced average evolution speed, accompanied by a pronounced excitonic population imbalance (the difference of remnant excitonic populations between the $K$ and $K'$ valleys).
The practical application of the two optimal schemes in system fidelity and valley polarization as well as the effect of magnetic field induced valley Zeeman splitting is also discussed.
Our results facilitate the practical application of the QSL theory in TMDC field, and stimulate valley quantum information probing more magneto-optical features.

The rest of the paper is organized as follows:
In Sec.~\ref{sec:theo}, we show our theoretical framework describing both intra- and intervalley scatterings of excitons in monolayer WSe$_2$, which are essential for determining the QSL time and evolution speed of the valley system.
In this context, we derive the differential equations of density matrix of the system, and then propose two optimal protocols of valley dynamics based on the QSL theory.
In Sec.~\ref{sec:resu}, we investigate the associated controls of valley dynamics for reducing evolution time and for accelerating the evolution speed, by means of the optical excitation and magnetic field.
Also, we emphasize their practical applications in system fidelity and valley polarization.
Finally, we summarize our main conclusions in Sec.~\ref{sec:conc}.

\section{Theoretical framework}
\label{sec:theo}

We begin by considering monolayer WSe$_2$ [Fig.~\ref{figure1}(a)] as the physical system to study the associated optimal proposal of valley dynamics, which can be extended to other W-based and even Mo-based TMDC materials.
To this end, we first introduce the intra- and intervalley channels of excitons, and then give a dynamical description of the valley system interacted with a laser field.
Based on the geometric QSL theory, we introduce two types of optimal protocols of valley dynamical evolution.

\subsection{Intravalley channels: selective excitation and radiative recombination}

For the intravalley channels, the conduction band minima and valence band maxima are both located at the degenerate $K$ and $K'$ points at the corners of Brillouin zone [Fig.~\ref{figure1}(b)].
As the $K$ and $K'$ valleys are related to each other by time-reversal symmetry, there is a one-to-one correspondence between two valleys for either the optical excitation or the recombination process.
Proceeding from the orbital symmetry, the threefold rotation ($C_3$) splits the $d$ orbitals of transition metal atoms into three groups: $A_1(d_{z^2})$, $E(d_{xy}, d_{x^2-y^2})$, $E'(d_{xz}, d_{yz})$.
The first-principles calculation indicates that the wave functions at conduction ($|\uppsi_c\rangle$) and valence ($|\uppsi_v\rangle$) band edges are $|\uppsi_c\rangle=|d_{z^2}\rangle$ and $|\uppsi_v^{\xi}\rangle=\frac{1}{\sqrt{2}}(|d_{x^2-y^2}\rangle+i\xi|d_{xy}\rangle)$~\cite{PhysRevLett.108.196802,cao2012}, with $\xi=\pm1$ the valley index.
Further, the $C_3$ rotational operation of TMDCs requires the symmetry adapted basis functions satisfy $C_3|\uppsi_a\rangle=\exp(-i\frac{2\pi}{3}m_a)|\uppsi_a\rangle~(a=c,v)$,
with $m_a$ the magnetic quantum numbers for conduction and valence bands extrema.
Then considering the transition matrix element $W_{cv}=\langle\uppsi_c|\mathcal{P}_\pm|\uppsi_v\rangle$, with $\mathcal{P}_\pm=\mathcal{P}_x\pm i\mathcal{P}_y$ the electric dipole moment, the chiral optical selectivity can be deduced, corresponding to the absorption of right- ($\sigma_+$) and left-handed ($\sigma_-$) photons at the $K$ and $K'$ valleys, respectively.

Regarding the radiative recombination of excitons, it is one important process that leads to the energy loss and decoherence, which can be described by the Fermi's golden rule.
The exciton emits one photon with momentum $\mathbf{q}$ and light mode $\uplambda$, from an excited state $|\Psi_\xi(\mathbf{Q}),0\rangle$ ($\xi=K,K'$) to the ground state $|G,1_{\uplambda\mathbf{q}}\rangle$.
In this process, the decay rate of an excitonic state $\xi$ with center-of-mass momentum $\mathbf{Q}$ is given by $\gamma_\xi(\mathbf{Q})=\frac{1}{\uptau_{\xi}(\mathbf{Q})}=\frac{2\pi}{\hbar}\sum_{\uplambda\mathbf{q}}|\langle G,1_{\uplambda\mathbf{q}}|H_{\rm{LM}}|\Psi_{\xi}(\mathbf{Q}),0\rangle|^2
\delta[E_{\xi}(\mathbf{Q})-E_{\uplambda\mathbf{q}}]$~\cite{PhysRevLett.95.247402}, with $H_{\rm{LM}}$ the light-matter interaction Hamiltonian in the dipole approximation, $E_{\xi}(\mathbf{Q})$ and $E_{\uplambda\mathbf{q}}$ the energies of exciton and photon, respectively.
Here the excitonic state $|\Psi_{\xi}(\mathbf{Q})\rangle$ can be unfolded as a superposition of hole (with crystal momentum $\mathbf{k}$) and electron (with crystal momentum $\mathbf{k}+\mathbf{Q}$) states from band pairs ($v$, $c$) in the reciprocal space, namely, $|\Psi_{\xi}(\mathbf{Q})\rangle=\sum_{c,v,\mathbf{k}}A^\xi_{c,v,\mathbf{k},\mathbf{Q}}|c,\mathbf{k}+\mathbf{Q}\rangle|v,\mathbf{k}\rangle$, with $A^\xi_{c,v,\mathbf{k},\mathbf{Q}}$ the Bethe-Salpeter equation (BSE) expansion coefficient.
Note that the decay rate at momentum $\mathbf{Q}=0$ reads $\gamma_\xi(0)=8\pi e^2E_{\xi}(0)\mu_\xi^2/\hbar^2cA_{uc}$~\cite{doi:10.1021/nl503799t}, with $\mu_\xi^2$ the square modulus of the BSE exciton transition dipole element, $c$ the speed of light, and $A_{uc}$ the area of the unit cell.
Then, considering the thermal effect in exciton recombination process, the radiative lifetime of an exciton in $\xi$ valley at temperature $T$ is $\uptau_\xi=\uptau_\xi^0[{3Mc^2k_{B}T}/{2E_{\xi}^2(0)}]$~\cite{doi:10.1021/nl503799t}, with $\uptau_\xi^0$ the recombination time near $T=0$, and $M$ the exciton mass.
Hence, the exciton lifetime increases linearly with elevated temperature, while inversely proportional to the square of excitonic energy $E_{\xi}(0)$ at $K$ and $K'$ valleys.

\subsection{Intervalley channels: electron-hole exchange and exciton-phonon interactions}

Since the intervalley scattering of excitons requires not only the spin-flip of electrons and holes, but also the momentum conservation, there are two dominated intervalley scattering mechanisms: electron-hole (e-h) exchange and exciton-phonon (ex-ph) interactions.
For the former, in the basis $\{|K,\mathbf{Q}\rangle,|K',\mathbf{Q}\rangle\}$, the long-range intervalley exchange interaction can be written as $H_{\rm{ex}}=J_{Q}[\cos(2\theta)\sigma_x+\sin(2\theta)\sigma_y]$~\cite{PhysRevB.89.205303,PhysRevB.47.15776,hao_moody2016,PhysRevLett.120.046402,PhysRevB.104.L121408}, with $\sigma_{x,y}$ the Pauli matrixs, $\mathbf{Q}$ the center-of-mass momentum, and $\theta$ the orientation angle.
Here the momentum-dependent exchange interaction reads $J_{Q}=\frac{\pi}{4}\alpha^2|\psi(0)|^2\sqrt{2T_Q/\rm{Ry}}$,
with $\rm{Ry}$ the Rydberg energy, $\alpha$ the effective fine structure, $T_Q$ the kinetic energy of center-of-mass motion, and $|\psi(0)|^2$ the probability that an electron and a hole spatially overlap respectively~\cite{yuliu2014}.
In monolayer WSe$_2$, we consider $\rm{Ry}=75.66$ meV, $\alpha=0.66$, and $|\psi(0)|=0.88$~\cite{hao_moody2016}, leading to a consequence of $J_{Q}=20~\rm{meV}\times\sqrt{T_Q/37.83}$.
Hence, the magnitude of e-h exchange interaction essentially scales linearly with the center-of-mass momentum of exciton, and direction depends on the orientation of exciton momentum.

Note that, both the magnitude and direction of $\mathbf{Q}$-dependent exchange interaction are random during the intervalley process, due to various scatterings from phonons, other excitons and defects, that provides an effective in-plane magnetic field driving valley pseudospin precession with different frequencies~\cite{PhysRevB.104.L121408,PhysRevB.107.035404}.
This is similar to the one that deals with the D’ykonov-Perel (DP) spin relaxation for 2D electron gases in the diffusive regimes~\cite{PhysRevB.90.245302}.
As a consequence, the valley pseudospin precession leads to an incoherent intervalley transfer [Fig.~\ref{figure1}(b)], which is manifested by a statistical average intervalley scattering time $\uptau_{\rm{eh}}$~\cite{PhysRevB.108.035419}.
However, in the presence of external magnetic field, the expectation value of pseudospin depends on the combined contributions of in-plane (e-h exchange interaction) and out-of-plane (external vertical magnetic field) components.
The external magnetic field-induced valley splitting suppresses the intervalley scattering related to exchange interaction~\cite{PhysRevB.108.035419}.
Thus, the exchange-related scattering rate reads $\Gamma_{\rm{eh}}=1/\uptau_{\rm{eh}}\times F(\Delta E)$, where $F(\Delta E)=\Gamma^2/(\Gamma^2+\Delta E^2)$~\cite{nanolett.8b01484,acs.jpcc.2c05113}, with $\Gamma$ the width parameter associated with exciton momentum relaxation~\cite{PhysRevB.50.14246}, and $\Delta E=E_{K'}-E_K$ the valley Zeeman splitting.

Regarding the ex-ph interaction, both the electron and hole confined in an exciton require one $K$-point phonon to satisfy zero center-of-mass momentum of the exciton before and after intervalley scatterings.
For this, the intervalley scattering rate for momentum conservation is proportional to the phonon occupation number, i.e., $\Gamma_{\rm{ph}}\propto\exp(-\langle\hbar\omega_{\rm{ph}}\rangle/k_BT)$~\cite{zeng2012valley}, with $\langle\hbar\omega_{\rm{ph}}\rangle=12$ meV the acoustic phonon energy near the $K$-point~\cite{PhysRevB.87.115418,PhysRevLett.127.157403,kioseoglou2016,carvalho2017}, closing to the acoustic phonon energy reported in TMDCs~\cite{PhysRevB.44.3955,zeng2012valley,Helmrich_2018}.
Note that the lifting of valley degeneracy due to magnetic field causes an asymmetric phonon-assisted relaxation process, which requires that excitonic scatterings from the valley of higher energy to the one of lower energy emitting an additional phonon ($\Gamma_H$), whereas absorbing a phonon occurs in the opposite process ($\Gamma_L$).
Consequently, the scattering rates are mediated by the Boltzmann factor and can be expressed as Miller form, namely, $\Gamma_{H}=\Gamma_{\rm{ph}}$ and $\Gamma_{L}=\Gamma_{\rm{ph}}\exp(-\Delta E/k_BT)$~\cite{doi:10.1063/1.5112823,PhysRev.120.745}.

Combining two part contributions from e-h exchange and ex-ph interactions, the total intervalley scattering rates can be written as $\Gamma_{u,d}=\Gamma_{\rm{eh}}+\Gamma_{H,L}$~\cite{PhysRevB.108.035419}.
Figure~\ref{figure1}(c) displays the intervalley scattering time $T_{u,d}=1/\Gamma_{u,d}$ as a function of magnetic field at low temperature.
We observe that the scattering time $T_u$ grows with magnetic field and rapidly reaches stability.
This is because the large Zeeman splitting regarding strong magnetic field fully quenches the exchange interaction-induced intervalley scattering, while phonon-assisted intervalley process from a higher valley in energy to lower valley is unaffected.
In contrast, the intervalley scattering from a lower valley in energy to higher valley is heavily suppressed due to the valley splitting, leading to a dramatic rise in scattering time $T_d$.
Furthermore, the intervalley scattering not only causes incoherent transfer of excitonic population from one valley to another, but also an additional exciton valley decoherence (i.e., pure dephasing), which is, an essential component of valley dynamics.
Considering the Maialle-Silva-Sham (MSS) mechanism, the random e-h exchange interaction induces an additional coherence loss, that is analogous to dephasing in conventional semiconductors and spin depolarization in germanium, also driven by intervalley scattering~\cite{hao_moody2016,PhysRevLett.111.257204,yangsinitsyn2015}.
Also, optical two-dimensional coherent spectroscopy reveals that the ex-ph interaction contributes significantly to pure dephasing~\cite{moody2015intrinsic}, which should be taken into account.

\subsection{Valley dynamical evolution: model Hamiltonian and master equation}

In order to control the valley dynamical evolution in optical devices, we mainly consider a two-valley system, which interacts with a laser field [Fig.~\ref{figure1}(d)].
Further, we consider the Zeeman shifts of energy bands caused by an external vertical magnetic field, which induces an asymmetric intervalley scattering from $K$ ($K'$) valley to $K'$ ($K$) valley with the scattering rate $\Gamma_d$ ($\Gamma_u$).
This allows us to construct a model with three-level states, namely, excitons in the $K$ ($|K\rangle$) and $K'$ ($|K\rangle$) valleys and the ground state $(|0\rangle)$~\cite{PhysRevB.81.075322}.

Incorporate the intra-and intervalley scattering channels into the master equation of Lindblad form, then we employ the density matrix to illustrate the valley dynamics.
The total dynamical evolution of system is composed of four parts: unitary process, intervalley scattering, intravalley exciton recombination and pure dephasing, satisfies
\begin{equation}
\label{eqa11}
\frac{d}{dt}\rho_t=L_0(\rho_t)+L_f(\rho_t)+L_r(\rho_t)+L_p(\rho_t).
\end{equation}
For the part of unitary evolution, it is driven by the system Hamiltonian and can be described by the Liouville operator $L_0(\rho_t)=-i[H,\rho_t]$.
In the basis $\{|K\rangle,|K'\rangle,|0\rangle\}$, the system Hamiltonian reads, $H=H_{\rm{ex}}+H_{\rm{In}}$, comprising contributions from the valley ($H_{\rm{ex}}$), and the interaction term ($H_{\rm{In}}$).
Here the first term,
\begin{equation}
\label{eqa12}
\begin{split}
H_{\rm{ex}}=\sum_{\xi}E_{\xi}c_{\xi}^{\dag}c_{\xi},~\xi=K,K',
\end{split}
\end{equation}
describes the exciton energies at separate $K$ and $K'$ valleys.
In the presence of magnetically-dependent Zeeman effect, the exciton energy is written as $E_{\xi}=E_0+\xi\mu_BgB/2$, with $E_0$ the energy at zero field, $\xi$ the valley index ($\xi=1$ for $K$ and $\xi=-1$ for $K'$ valley), $\mu_B$ the Bohr magneton, and $g$ the Lande factor for bright exciton in monolayer WSe$_2$.
$c_{\xi}^{\dag}(c_{\xi})$ stands for the exciton creation (annihilation) operator of the $\xi$ valley.
The relationship between magnetic field and energy levels of two valleys is displayed in Fig.~\ref{figure1}(e).

The second term in the form of rotating wave approximation~\cite{Breuerbook}, giving rise to Rabi oscillations between the ground state ($|0\rangle$) and excitonic states ($|K\rangle$ and $|K'\rangle$), is expressed as~\cite{PhysRevB.69.125342,PhysRevB.107.035404,PhysRevB.98.075423}
\begin{equation}
\label{eqa13}
\begin{split}
H_{\rm{In}}=g_Ke^{i\omega_Rt}\sigma_{K-}+g_{K'}e^{i\omega_Lt}\sigma_{K'-}+\rm{H.c.},
\end{split}
\end{equation}
where $\omega_R$ and $\omega_L$ are the frequencies for right ($\sigma_+$) and left ($\sigma_-$) circularly polarized excitation modes, respectively.
We define $\sigma_{\xi-}=|0\rangle\langle\xi|$ as the lowering operators and $\rm{H.c.}$ means the Hermitian conjugate.
The parameter $g_{\xi}=\langle 0|\vec{\mu}\cdot\vec{E}|\xi\rangle/2$ is the dipole coupling strength for $\sigma_{\pm}$ excitation, with $\vec{\mu}$ the electric dipole moment and $\vec{E}$ the amplitude of laser field.
Without loss of generality, we consider $g_{\xi}$ is real with $g_{\xi}=g_{\xi}^*$.
To eliminate the time-dependence induced by light-matter interaction in the Eq.~(\ref{eqa13}), we apply an unitary transformation on the initial system Hamiltonian $H$, namely, $\mathcal{H}=UHU^{\dag}+i\hbar\dot{U}U^{\dag}$~\cite{PhysRevB.69.125342}, where the transformation matrix
\begin{equation}
\label{eqa14}
\begin{split}
U=e^{i\frac{\omega_{RL}}{2}t}|K\rangle\langle K|+e^{-i\frac{\omega_{RL}}{2}t}|K'\rangle\langle K'|+e^{-i\frac{\omega_{R}+\omega_{L}}{2}t}|0\rangle\langle 0|,
\end{split}
\end{equation}
with $\omega_{RL}=\omega_{R}-\omega_{L}$.
Hence the transformed system Hamiltonian can be written as
\begin{equation}
\label{eqa15}
\begin{split}
\mathcal{H}=\sum_{\xi=K,K'}\big[\big(\Delta_\xi+\frac{1}{2}\xi\mu_BgB\big)c_{\xi}^{\dag}c_{\xi}+g_\xi(\sigma_{\xi-}+\rm{H.c.})\big],
\end{split}
\end{equation}
with $\Delta_K=E_0-\hbar\omega_R$ and $\Delta_{K'}=E_0-\hbar\omega_L$ the exciton-laser field detuning at zero magnetic field.
We define the symmetric and asymmetric excitations as $\Delta_K=\Delta_{K'}$ ($\omega_R=\omega_L$) and $\Delta_K\not=\Delta_{K'}$ ($\omega_R\not=\omega_L$), respectively.
Further, in the presence of magnetic field, which gives rise to the valley Zeeman splitting $\Delta E$, we introduce the detuning difference as $\Delta_d=\left|\Delta_K-\Delta_{K'}-\Delta E\right|$ to manifest the effect of magnetic field on the exciton-laser field detuning.
Note that in the absence of magnetic field, the detuning difference depends only on the zero-field detuning $\Delta_\xi$.
Substituting Eq.~(\ref{eqa15}) into the Liouville equation, we obtain the unitary evolution part of valley dynamics.

The intervalley scattering caused by the e-h exchange and ex-ph interactions, leads to an incoherent transfer of excitonic population from one valley to another~\cite{PhysRevB.92.235425,qiu2019roomtemperature}, described by $\dot{\rho}^K=-\Gamma_d\rho^K+\Gamma_u\rho^{K'}$, and $\dot{\rho}^{K'}=\Gamma_d\rho^K-\Gamma_u\rho^{K'}$, with $\rho^K$ and $\rho^{K'}$ the probabilities of finding exciton located in the $K$ and $K'$ valleys after recombination, respectively.
Also, the quantum coherence transfer will be limited by the intervalley scattering process~\cite{PhysRevA.103.043713}.
For the part of intravalley exciton recombination, the related description is provided by the following equation of Lindblad form $L_r(\rho_t)=\sum_{\xi=K,K'}\Gamma_{\xi}(\sigma_{\xi-}\rho_t\sigma_{\xi+}-\frac{1}{2}\{\sigma_{\xi+}\sigma_{\xi-},\rho_t\})$~\cite{foot1}, with $\Gamma_\xi=1/\uptau_{\xi}$ the temperature-dependent decay rate.
Note that in the presence of magnetic field, the opposite Zeeman shifts in $K$ and $K'$ valleys contribute to the difference in $\Gamma_K$ and $\Gamma_{K'}$.
Beyond the loss of remnant excitonic population, the recombination also leads to a valley decoherence, which provides an upper bound for coherence time~\cite{hao_moody2016}.
Further, the e-h exchange and ex-ph interactions induce the pure dephasing of exciton valley coherence~\cite{hao_moody2016,PhysRevB.108.035419,yangsinitsyn2015}, which can be expressed by this Lindblad operator $L_p(\rho_t)=\sum_{n=K,K',0}\upgamma(\sigma_{nn}\rho_t\sigma_{nn}^{\dag}-\frac{1}{2}\{\sigma_{n}\sigma_{n}^{\dag},\rho_t\})$~\cite{PhysRevB.108.035419}, with $\upgamma$ referring to the pure dephasing rate.

Thus, in terms of the above dynamical operators, the evolution of motion for the density matrix can be written as a set of coupled differential equations
\begin{equation}
\label{eqa19}
\begin{split}
\dot{\rho}^{K}&=ig_K(\rho^{K0}-\rho^{0K})-(\Gamma_K+\Gamma_d)\rho^{K}+\Gamma_u\rho^{K'},\\
\dot{\rho}^{K'}&=ig_{K'}(\rho^{K'0}-\rho^{0K'})+\Gamma_d\rho^{K}-(\Gamma_{K'}-\Gamma_u)\rho^{K'},\\
\dot{\rho}^{KK'}&=i\big[(\Delta_{K'}-\Delta_K-\Delta E)\rho^{KK'}-g_K\rho^{0K'}+g_{K'}\rho^{K0}\big]\\
&-\Gamma_r\rho^{KK'}-\upgamma\rho^{KK'},\\
\dot{\rho}^{K0}&=i\big[-(\Delta_K-\frac{\Delta E}{2})\rho^{K0}+g_K(\rho^K-\rho^0)+g_{K'}\rho^{KK'}\big]\\
&-\frac{\Gamma_K}{2}\rho^{K0}-\upgamma\rho^{K0},\\
\dot{\rho}^{K'0}&=i\big[-(\Delta_{K'}+\frac{\Delta E}{2})\rho^{K'0}+g_{K'}(\rho^{K'}-\rho^0)+g_K\rho^{K'K}\big]\\
&-\frac{\Gamma_{K'}}{2}\rho^{K'0}-\upgamma\rho^{K'0},
\end{split}
\end{equation}
with the parameter $\Gamma_r=(\Gamma_K+\Gamma_{K'})/2$.
Other matrix elements are defined as $\rho^0=1-\sum_\xi\rho^\xi$, $\rho^{K'K}=(\rho^{KK'})^*$ and $\rho^{0\xi}=(\rho^{\xi0})^*$.
The diagonal terms $\rho^{K(K')}$ and $\rho^0$ denote populations of excitons locating three energy levels, respectively, while off-diagonal terms $\rho^{ij}(i\not=j)$ describe the coherences between three occupied states (i.e., $|K\rangle$, $|K'\rangle$ and $|0\rangle$).
Here the coherence intensity characterizing the degree of valley coherence can be defined as $\mathcal{C}(\rho)=|\rho^{KK'}|+|\rho^{K'K}|$~\cite{Lan_2022,PhysRevB.108.035419}, with $0\leq\mathcal{C}(\rho)\leq1$.
Also, the coherence time $\uptau_C$ can be interpreted as the time over which $\mathcal{C}(\rho)$ essentially vanishes.

\subsection{Optimal protocol of valley dynamics: geometric quantum speed limit}
\label{sec:qslw}
In this subsection, we first outline the geometric QSL theory, and introduce both the minimum evolution time (i.e., the QSL time) and average evolution speed of valley dynamics, from which, we further illustrate two types of optimal protocols of valley dynamics.

The QSL sets the lower bound on the evolution time between two distinguishable states of a quantum system~\cite{Hornedal2023,PhysRevX.6.021031}, that has manifold applications in quantum coherence~\cite{Mohan_2022}, quantum resource theories~\cite{Campaioli_2022}, optimal control~\cite{Deffner_2014,PhysRevLett.111.260501,PhysRevLett.103.240501}, and quantum thermodynamics~\cite{Binder_2015,PhysRevX.13.011013}, quantum battery~\cite{PhysRevA.102.060201,PhysRevLett.118.150601}, among other fields.
For an open dynamical process, geometric approach is a typical tool to study the QSL problems, focusing on the principle that the geodesic path $\mathcal{D}$ is the shortest one among all dynamical evolutions between two distinguishable quantum states $|\uppsi_0\rangle$ and $|\uppsi_\uptau\rangle$ [Fig.~\ref{figure1}(f)]~\cite{PhysRevX.6.021031}.
Physically, the geometric QSL time for a quantum system evolving from an initial state $\rho_0$ to a final state $\rho_\uptau$ can be
derived from the inequality between the lengths of the geodesic and actual path~\cite{PhysRevLett.110.050402},
\begin{equation}
\label{eqb1}
\begin{split}
 \mathcal{D}(\rho_0,\rho_\uptau)\leq\sum_{i=1}^{n}\mathcal{D}(\rho_{(i-1)\Delta t},\rho_{i\Delta t})=\mathcal{L}(\rho_0,\rho_\uptau),
\end{split}
\end{equation}
where the actual evolution time $\uptau$ is divided into $n$ infinitesimal time $\Delta t=\uptau/n$ with $n\rightarrow+\infty$.
The saturable case with equality holds if the quantum system evolves from the initial state $\rho_{0}$ to final state $\rho_{\uptau}$ always along the geodesic.
Further, the instantaneous speed along the actual path can be expressed as $v(t)=\frac{d}{dt}\mathcal{L}(\rho_0,\rho_t)=\lim_{\Delta t\rightarrow0}\frac{\mathcal{D}(\rho_{t},\rho_{t+\Delta t})}{\Delta t}$~\cite{Lan_2022,PhysRevA.103.022210},
with $\lim_{\Delta t\rightarrow0}\mathcal{D}(\rho_{t},\rho_{t+\Delta t})=\lim_{\Delta t\rightarrow0}[\mathcal{L}(\rho_0,\rho_{t+\Delta t})-\mathcal{L}(\rho_0,\rho_t)]$.
Based on this, the generalized QSL time associated with the time-averaged speed of the actual evolution path of valley dynamics is $\uptau_{\rm{m}}=\frac{\mathcal{D}(\rho_0,\rho_\uptau)}{\bar{v}(\uptau)}$,
with $\bar{v}(\uptau)=(1/\uptau)\int_{0}^{\uptau}v(t)dt$ denoting the time average speed.

The geometric description of geodesic is crucial for deriving the QSL time in an open dynamical process~\cite{PhysRevX.6.021031}.
There are a class of QSLs that can be investigated based on different geometric metric, such as Bures angle~\cite{Lan_2022,PhysRevLett.110.050402,PhysRevLett.111.010402}, trace distance~\cite{PhysRevA.95.052104}, relative purity~\cite{PhysRevA.98.042132,PhysRevLett.120.060409}, quantum Fisher information~\cite{PhysRevLett.110.050402}, and Wigner-Yanase information~\cite{PhysRevX.6.021031,pires2023}.
Hereafter, we restrict our analysis to the geodesic length characterized by the Euclidean distance, given by $\mathcal{D}(\rho_0,\rho_\uptau)=\|\rho_\uptau-\rho_0\|_{\rm{hs}}$, with the Hilbert-Schmidt norm $\|X\|_{\rm{hs}}=\sqrt{\text{tr}(X^{\dag}X)}$~\cite{Bhatia1996MatrixA}.
Hence, the QSL time regarding the Euclidean distance reads~\cite{2019tightrobust}
\begin{equation}
\label{eqa21}
\begin{split}
\uptau_{\rm{m}}=\frac{\|\rho_\uptau-\rho_0\|_{\rm{hs}}}
{\bar{v}(\uptau)},
\end{split}
\end{equation}
with the time-average speed of dynamical evolution~\cite{2019tightrobust,PhysRevLett.111.010402,PhysRevA.95.052104,PhysRevA.98.042132}
\begin{equation}
\label{eqb2}
\begin{split}
\bar{v}(\uptau)=\overline{\|\dot{\rho}_t\|}_{\rm{hs}}=(1/\uptau)\int_{0}^{\uptau}dt\|\dot{\rho}_t\|_{\rm{hs}}.
\end{split}
\end{equation}
Further, employing the density matrix of system obtained from Eq.~(\ref{eqa19}), we obtain the geodesic length and average evolution speed of valley dynamics from an initial state $\rho_0$ to a final state $\rho_\uptau$
\begin{equation}
\label{eqa22}
\begin{split}
&\mathcal{D}(\rho_0,\rho_\uptau)=\sqrt{\sum_{n}\big(\rho_0^n-\rho_\uptau^n\big)^2
+2\sum_{i\not=j}S^{ij}_\uptau S^{ji}_\uptau},\\
&\bar v(\uptau)=\frac{1}{\uptau}\int_0^\uptau dt\sqrt{\sum_{n}(\dot{\rho}_t^n)^2+2\sum_{i\not=j}\dot{\rho}^{ij}_t\dot{\rho}^{ji}_t},
\end{split}
\end{equation}
with the parameter $S^{ij}_\uptau=\rho^{ij}_0-\rho^{ij}_\uptau(i\not=j)$ and $n=K,K',0$.
Further, the actual evolution length reads $\mathcal{L}(\rho_0,\rho_\uptau)=\bar v(\uptau)\uptau$.
This evolution speed in Eq.~(\ref{eqa22}) describes the overall behavior of valley dynamics, involving in the exciton generation, intravalley recombination, intervalley scattering, together with valley coherence loss.

Optimal control theory aims to find a feasible control pattern, to optimize a particular \emph{performance measure}~\cite{Deffner_2014,1959Classical}.
In the paradigm of valley dynamics, we mainly consider the following two \emph{performance measures} to be optimized: (i) the actual evolution time of evolving to the target state ($\uptau$), and (ii) the average evolution speed over a period of time [$\bar v(\uptau)$].
For the former, we use the ratio of the QSL time over the actual time (i.e., $\uptau_{\rm{m}}/\uptau$) to investigate the optimal scheme, which indicates how far the valley dynamical evolution is from the selected geodesic path~\cite{PhysRevX.6.021031,Lan_2022}.
When the actual evolution path and the geodesic path coincide, the ratio is equal to one~\cite{PhysRevLett.123.180403,PhysRevLett.127.100404}, which is the optimal situation.
This allows the valley system to evolve along a path that takes the minimum time.
For the latter, the optimal control calls for a faster evolution speed $\bar v(\uptau)$ over a period of time ($0\sim\uptau$), which is only associated with the actual evolution path $\mathcal{L}$.
Note that there is significant difference between two types of optimal protocols.
The reduction in actual evolution time largely requires that the valley system evolves along a shorter path.
However, accelerating the valley dynamics would not necessarily require the system evolution path to converge towards the geodesic path.
Sometimes, the valley dynamics may have a faster speed in a longer evolution path.

\section{Results and discussions}
\label{sec:resu}

Regarding the quantum information processing using valley degree of freedom, the first important step is the generation of a superposition of two valleys as $|\uppsi_0\rangle=1/\sqrt{2}(|K\rangle+|K'\rangle)$, which is optically achieved by linearly polarized excitation, and can be detected by a strongly linearly polarized emission~\cite{jones_yu2013,PhysRevLett.115.117401,PhysRevB.90.075413}.
For practical considerations in monolayer TMDCs, the energy of optical excitation may not be exactly the same as the energy of electronic transitions~\cite{PhysRevLett.115.117401}, suggesting that it is feasible to tune the valley dynamics through the exciton-field detuning.
For our simulation, we use the parameters: exciton energy at zero-field $E_0=1.7~\rm{eV}$~\cite{aivazianmagnetic2015}, Lande factor $g=-4.25$~\cite{tepliakov2020}, width parameter $\Gamma=10^{-4}$ eV~\cite{nanolett.8b01484},
exchange induced scattering time $\uptau_{\rm{eh}}=50$ ps~\cite{nanolett.8b01484,PhysRevB.97.115425}, phonon-assisted scattering time $\uptau_{\rm{ph}}=1/\Gamma_{\rm{ph}}=50$ ps~\cite{Baranowski2017}, radiative recombination time at zero-field $\uptau_\xi(=4~\rm{K})=3.8~\rm{ps}$~\cite{doi:10.1021/nl503799t} and pure dephasing rate $\upgamma=5$ ps$^{-1}$~\cite{hao_moody2016}.
Also we consider a low-intensity incident pulse $g_\xi=1~\rm{meV}$~\cite{PhysRevB.81.075322,PhysRevA.68.012310}, in which the optical coupling strength is significantly smaller than the exciton energy.
These are typical values for bright excitons in monolayer WSe$_2$.
In the following, we will demonstrate the control of valley dynamical evolution, by adjusting the detuning $\Delta_{\xi}$ and external magnetic field.

\subsection{Valley dynamical optimization: reducing the evolution time}

\begin{figure}
\includegraphics[width=\linewidth]{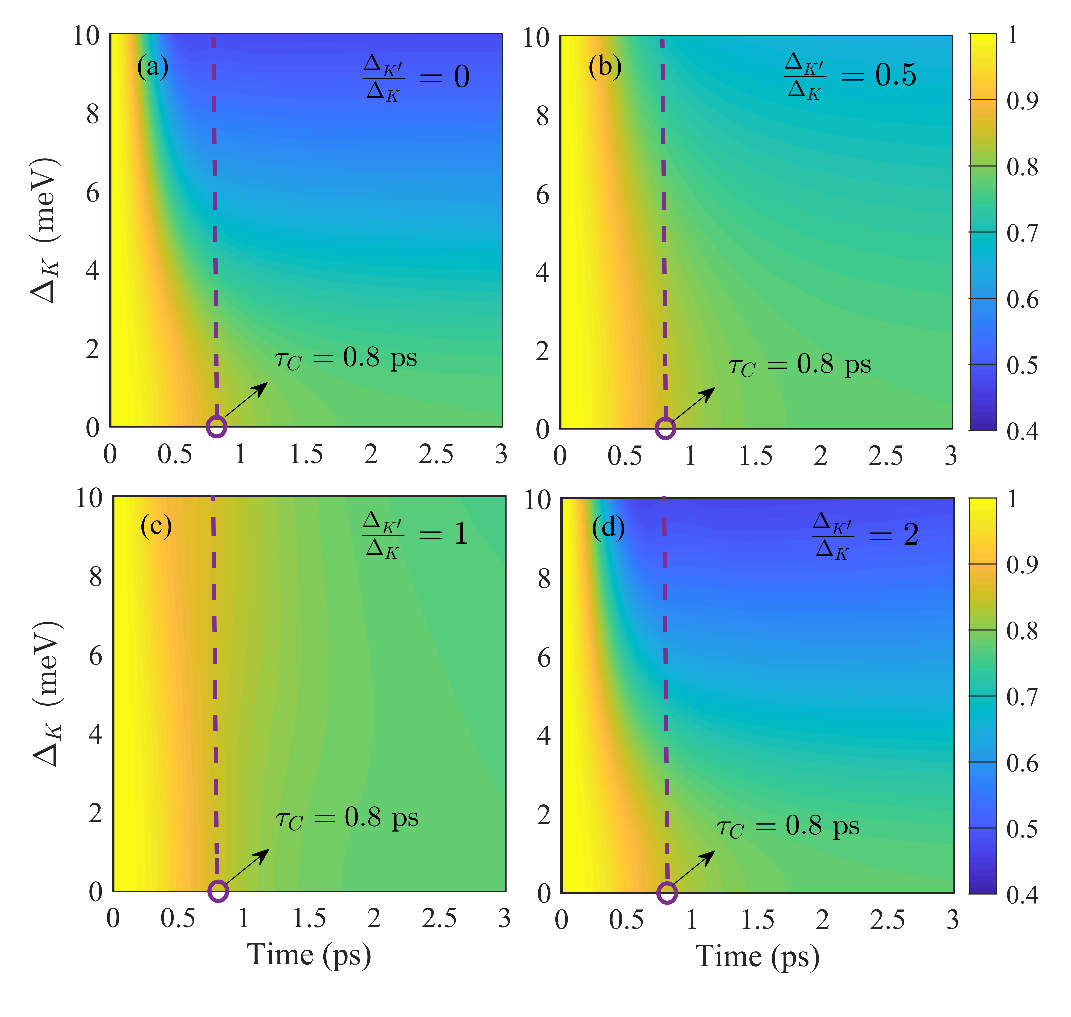}
\caption{The ratio $\uptau_{\rm{m}}/\uptau$ as a function of the actual evolution time $\uptau$ and detuning $\Delta_K$ at $B=0$, for four ratios of $\Delta_{K'}$ over $\Delta_K$, with $\Delta_{K'}/\Delta_K=0$ (a), $\Delta_{K'}/\Delta_K=0.5$ (b), $\Delta_{K'}/\Delta_K=1$ (c) and $\Delta_{K'}/\Delta_K=2$ (d).
In (a) and (d), the detuning difference $\Delta_d=\Delta_K$.
In (b), the detuning difference $\Delta_d=0.5\Delta_K$.
In (c), the detuning difference $\Delta_d=0$.
The purple circles mark the valley coherence time $\uptau_C\approx0.8$ ps.
The left side of the purple dashed line refers to the coherent region ($\uptau\textless0.8$ ps), while the right side is for the incoherent region ($\uptau\textgreater0.8$ ps).}
\label{figure2}
\end{figure}

We investigate the mechanism for reducing the evolution time of valley dynamics by means of the detuning $\Delta_{\xi}$ and external magnetic field.
Our aim is to find the optimal situation in which the valley dynamical evolution time clinches tightest to the QSL time.
For this, we first show the ratio $\uptau_{\rm{m}}/\uptau$ as a function of the actual evolution time $\uptau$ and detuning $\Delta_K$ for four ratios of $\Delta_{K'}$ over $\Delta_K$ [Fig.~\ref{figure2}].
To simultaneously exhibit the coherence effect of valley dynamics, we obtain the valley coherence time $\uptau_C\approx0.8$ ps (see the purple circles in Fig.~\ref{figure2}), which is in reasonable agreement with the experimental results in two-dimensional coherent spectroscopy~\cite{hao_moody2016,PhysRevLett.117.187401}.
As the evolution time exceeds 0.8 ps, there is essentially no valley coherence behavior in the dynamical process, with $\mathcal{C}(\rho)\approx0$.
Overall, we find that the ratio $\uptau_{\rm{m}}/\uptau$ drops quickly with time after a period of stabilization and eventually reaches the minimum value.
Larger ratios emerge in the coherent region ($\uptau\textless0.8$ ps), suggesting that valley coherence plays a crucial role in driving the actual evolution path to converge to the geodesic path.
Especially, when using a symmetric excitation with $\Delta_K=\Delta_{K'}$ [Fig.~\ref{figure2}(c)], the ratio $\uptau_{\rm{m}}/\uptau$ remains stable as changing the detuning $\Delta_K$, referring to the system evolving essentially along the geodesic path in the coherent region.
However, when using an asymmetric excitation with $\Delta_K\not=\Delta_{K'}$ [Figs.~\ref{figure2}(a),~\ref{figure2}(b) and~\ref{figure2}(d)], the ratio $\uptau_{\rm{m}}/\uptau$ decreases significantly in the coherent region with the increase of detuning $\Delta_K$, which indicates a deviation between the actual evolution path and the geodesic path.

\begin{figure}
\includegraphics[width=\linewidth]{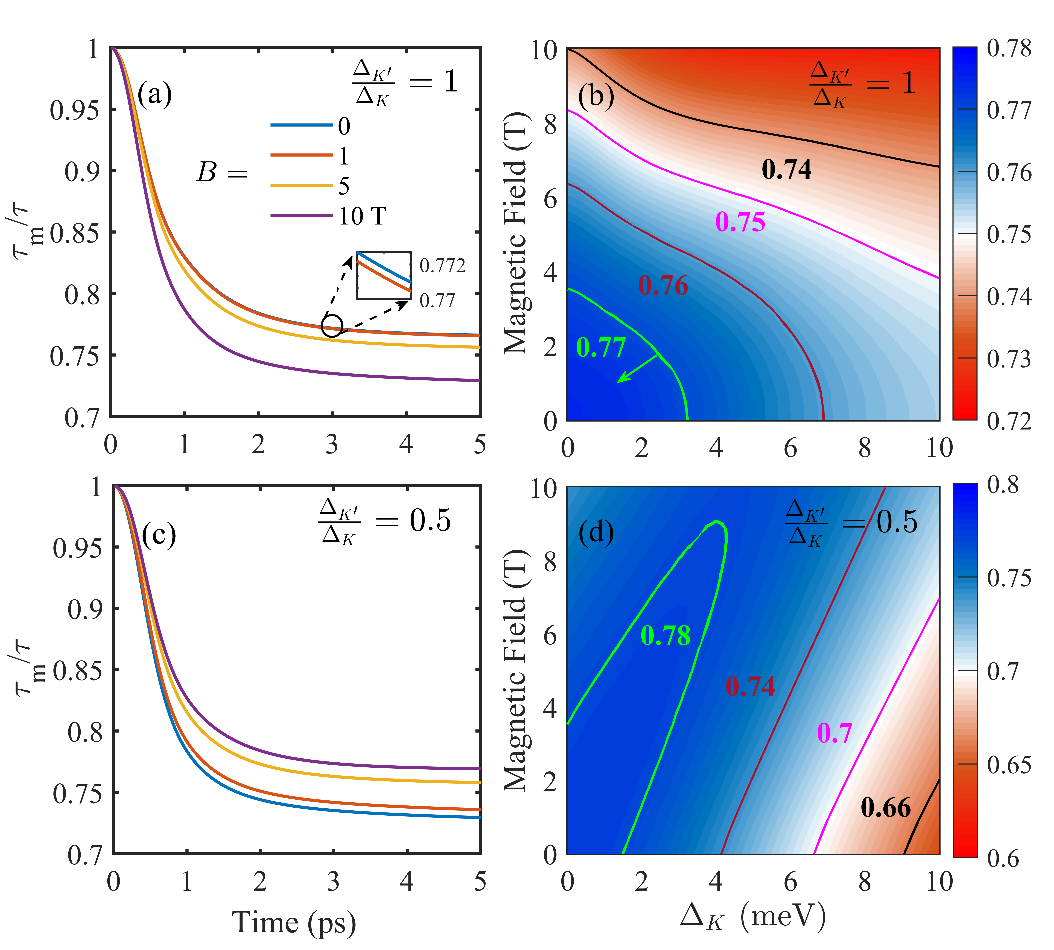}
\caption{Time evolutions of ratio $\uptau_{\rm{m}}/\uptau$ for different magnetic fields with symmetric excitation $\Delta_K=\Delta_{K'}$ (a) and asymmetric excitation $\Delta_K\not=\Delta_{K'}$ (c).
The ratio $\uptau_{\rm{m}}/\uptau$ as a function of detuning $\Delta_K$ and magnetic field with symmetric excitation $\Delta_K=\Delta_{K'}$ (b) and asymmetric excitation $\Delta_K\not=\Delta_{K'}$ (d).
Several values of contour lines of ratio $\uptau_{\rm{m}}/\uptau$ are also shown.
In (a) and (c), we choose the detuning $\Delta_K=5$ meV.
In (b) and (d), we consider the actual evolution time $\uptau=10$ ps.}
\label{figure3}
\end{figure}

As time goes on, the ratio $\uptau_{\rm{m}}/\uptau$ decreases further in the incoherent region ($\uptau\textgreater0.8$ ps).
This indicates that the actual evolution path further deviates from the geodesic path in the absence of valley coherence.
Also, we find that the ratio $\uptau_{\rm{m}}/\uptau$ in Fig.~\ref{figure2}(a) exhibits the identical behavior with that in Fig.~\ref{figure2}(d), in which, the ratio at the final state ($\uptau=3$ ps) can be dropped to 0.4.
This is attributed to the fact that the dynamical behavior of ratio $\uptau_{\rm{m}}/\uptau$ is mainly dominated by the detuning difference $\Delta_d$.
That is, in Figs.~\ref{figure2}(a) and~\ref{figure2}(d), the detuning differences are the same, with $\Delta_d=\Delta_K$, though the ratios of $\Delta_{K'}$ over $\Delta_K$ are not the same.
Remarkable, when the detuning difference $\Delta_d=0$ [Fig.~\ref{figure2}(c)], there is an optimal valley dynamical evolution, where the system evolves along a path closest to the geodesic path, accompanied with the largest ratio $\uptau_{\rm{m}}/\uptau\approx0.75$ at the final state.
In this situation, the ratio $\uptau_{\rm{m}}/\uptau$ exhibits a favourable robustness to the detuning $\Delta_K$.
In practice, as long as symmetric excitation is satisfied, the valley dynamics can travel along a path that takes the least time.

We next investigate the control of valley dynamics in the presence of an external magnetic field.
Note that in this situation, the optical excitation mode and the external magnetic field together determine the magnitude of the detuning difference $\Delta_d$.
For completeness, we consider two types of optical excitation modes with symmetric excitation [Figs.~\ref{figure3}(a) and~\ref{figure3}(b)] and asymmetric excitation [Figs.~\ref{figure3}(c) and~\ref{figure3}(d)].
From Figs.~\ref{figure3}(a) and~\ref{figure3}(c), we find that the ratio $\uptau_{\rm{m}}/\uptau$ under symmetric excitation exhibit a contrary dependence on the magnetic field as compare to that under asymmetric excitation.
Specifically, for the symmetric excitation [Fig.~\ref{figure3}(a)], the ratio $\uptau_{\rm{m}}/\uptau$ is moderately suppressed as the magnetic field increases.
Note that, the dynamical behavior of ratio $\uptau_{\rm{m}}/\uptau$ at $B=0$ matches almost that at $B=1$ T, see the black circle in Fig.~\ref{figure3}(a).
In contrast, for the asymmetric excitation [Fig.~\ref{figure3}(c)], the magnetic effect leads to an enhancement in the ratio $\uptau_{\rm{m}}/\uptau$, which effectively pushes the evolution path to approach the geodesic path.
The opposite dependence of ratio $\uptau_{\rm{m}}/\uptau$ on magnetic field in Figs.~\ref{figure3}(a) and~\ref{figure3}(c) is attributed to the contrary contribution of valley splitting $\Delta E$ to the detuning difference $\Delta_d$ under symmetric and asymmetric excitations within the parameters considered.
That is, by considering the detuning $\Delta_K=5$ meV, the detuning difference $\Delta_d$ rises from 0 to 2.47 meV under symmetric excitation, with the application of an external magnetic field ranging from 0 to 10 T.
However, under asymmetric excitation, the detuning difference $\Delta_d$ decays from 2.5 meV to 0.04 meV.
The magnetic field can be used as an additional effective knob in reducing the evolution time.

Figures~\ref{figure3}(b) and~\ref{figure3}(d) show the synergistic effect of detuning and magnetic field on the ratio $\uptau_{\rm{m}}/\uptau$.
Clearly, for the symmetric excitation [Fig.~\ref{figure3}(b)], the magnetic effect always lowers the ratio $\uptau_{\rm{m}}/\uptau$ within the considered detuning range ($0\sim10$ meV).
For the asymmetric excitation [Fig.~\ref{figure3}(d)], the magnetic field has a weak influence on the ratio $\uptau_{\rm{m}}/\uptau$ when the detuning $\Delta_K$ less than 4 meV.
As the detuning $\Delta_K$ increases beyond 4 meV, the presence of magnetically-induced valley splitting $\Delta E$ contributes to the suppression of detuning difference $\Delta_d$, which effectively improves the ratio $\uptau_{\rm{m}}/\uptau$ at strong magnetic fields.
Furthermore, from the contour lines of Figs.~\ref{figure3}(b) and~\ref{figure3}(d), we find that larger ratios emerge in the region surrounded by the green contour line.
This is because the detuning difference $\Delta_d$ in this region approaches zero, being significantly smaller than those in other regions.
The synergistic effect of detuning and magnetic field on the valley dynamics opens up further possibility for reducing the evolution time.

\subsection{Valley dynamical optimization: enhancing the evolution speed}

\begin{figure}
\includegraphics[width=\linewidth]{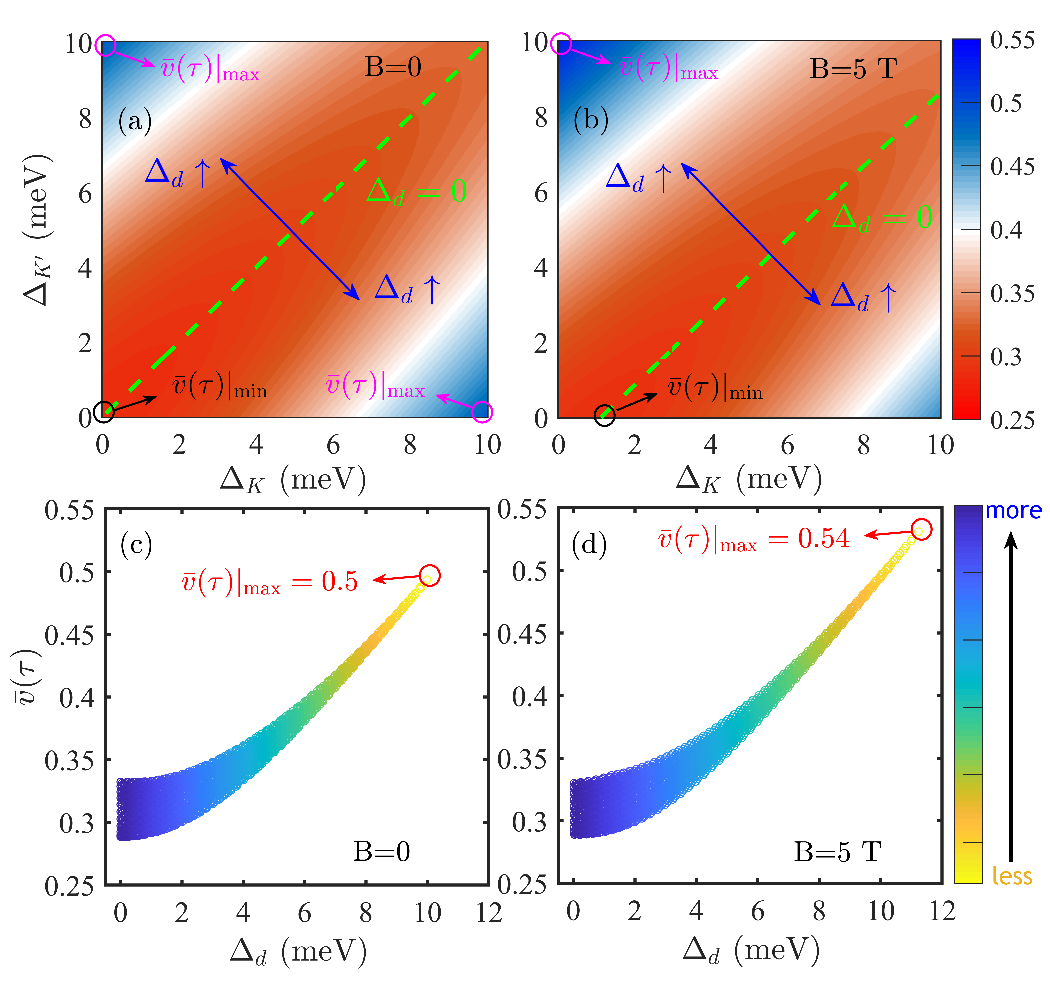}
\caption{The average evolution speed $\bar v(\uptau)$ of valley dynamics from $\uptau=0$ to $\uptau=4$ ps as a function of detunings $\Delta_K$ and $\Delta_{K'}$ at $B=0$ (a) and $B=5$ T (b).
The green dashed lines denote average evolution speed $\bar v(\uptau)$ when the detuning difference $\Delta_d=0$.
The pink (black) circles in (a) and (b) mark the minimum (maximum) evolution speed $\bar v(\uptau)|_{\min(\max)}$.
Blue double arrows in (a) and (b) denote the orientations where the detuning difference $\Delta_d$ increases.
The scatterplots of average evolution speed $\bar v(\uptau)$ with respect to the detuning difference $\Delta_d$ at $B=0$ (c) and $B=5$ T (d).
The red circles in (c) and (d) mark the maximum evolution speed $\bar v(\uptau)|_{\max}$.}
\label{figure4}
\end{figure}

We show the average evolution speed $\bar v(\uptau)$ of valley dynamics as a function of detunings $\Delta_K$ and $\Delta_{K'}$ at $B=0$ [Fig.~\ref{figure4}(a)] and $B=5$ T [Fig.~\ref{figure4}(b)].
In the absence of magnetic field [Fig.~\ref{figure4}(a)], the system evolves most slowly under symmetric excitation [see the green dashed line in Fig.~\ref{figure4}(a)], in which, the evolution speed $\bar v(\uptau)$ is essentially constant with changing the detuning $\Delta_\xi$.
Remarkable, we find that a large detuning difference $\Delta_d$ favors the the acceleration of valley dynamics.
As the detuning difference $\Delta_d$ grows from 0 to 10 meV [see the blue double arrows in Fig.~\ref{figure4}(a)], the evolution speed $\bar v(\uptau)$ can be boosted from 0.25 to 0.5, giving a noticeable rise, cf. the black and pink circles in Fig.~\ref{figure4}(a).
Note that, there are two positions where the detuning difference $\Delta_d$ is the largest, leading to the same evolution speed [see two pink circles in Fig.~\ref{figure4}(a)].
In the presence of magnetic field [Fig.~\ref{figure4}(b)], the green dashed line corresponding to the detuning difference $\Delta_d=0$ shifts to the right, which leads to the maximum evolution speed $\bar v(\uptau)|_{\max}$ occurring only at the position of $\Delta_{K'}-\Delta_{K}=10$ meV [see the pink circle in Fig.~\ref{figure4}(b)].
From the condition under which the maximum evolution speed arises in both Figs.~\ref{figure4}(a) and~\ref{figure4}(b), we emphasize that the asymmetric excitation is critical to contribute to a fast valley dynamical evolution.

We further explore the relationship between the evolution speed $\bar v(\uptau)$ and the detuning difference $\Delta_d$ at $B=0$ [Fig.~\ref{figure4}(c)] and $B=5$ T [Fig.~\ref{figure4}(d)].
As the detuning difference $\Delta_d$ increases from the minimum to the maximum, the color of data consequently transforms from blue to yellow, accompanied by a gradual narrowed speed range corresponding to the same detuning difference.
Overall, we find that the system is inclined to evolve rapidly in the case with a large detuning difference $\Delta_d$.
Meanwhile, since the detuning difference can be further increased by the valley splitting $\Delta E$, there is an enhancement in the maximum evolution speed at $B=5$ T versus the one at zero magnetic field, cf. the red circles in Fig.~\ref{figure4}(c) and~\ref{figure4}(d).

\begin{figure}
\includegraphics[width=\linewidth]{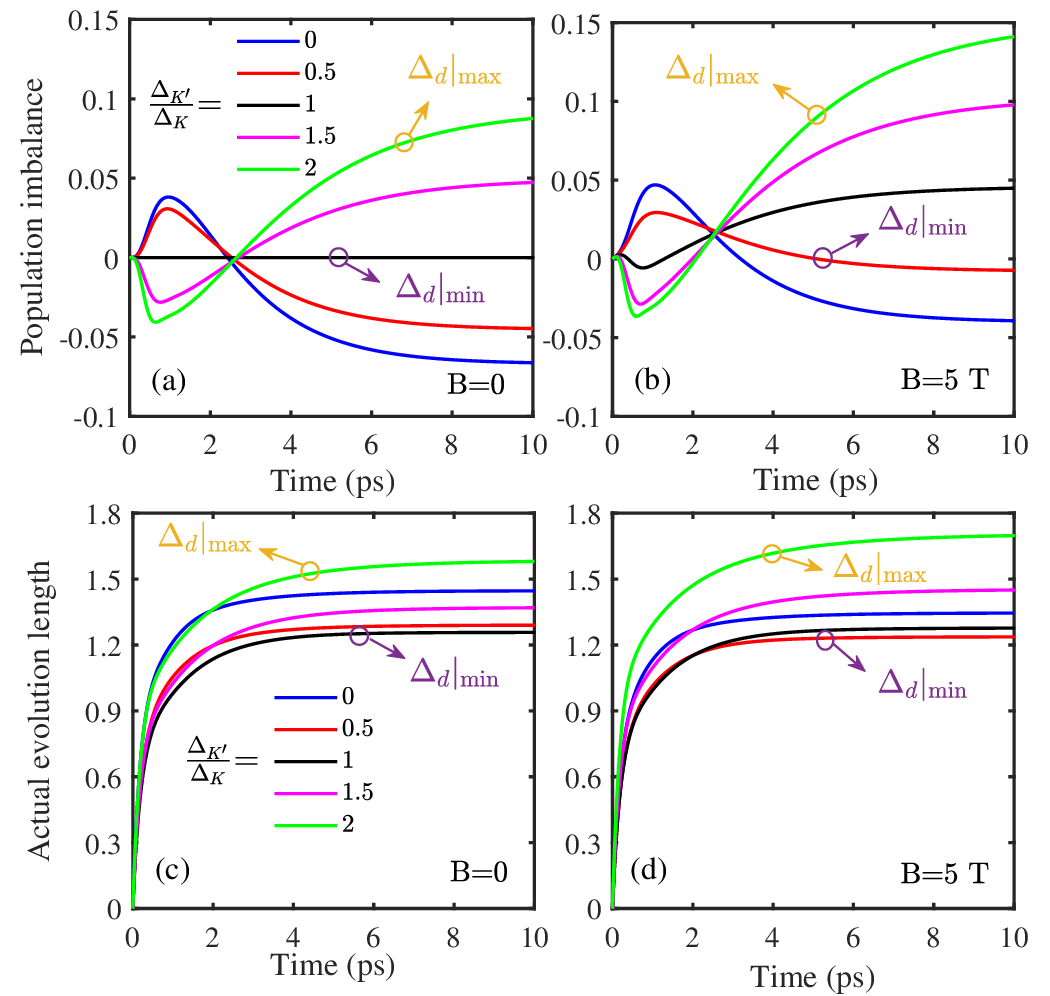}
\caption{(a) and (b) Time evolutions of population imbalance $\delta\rho=\rho^{K}-\rho^{K'}$, as well as (c) and (d) actual evolution length $\mathcal{L}(\rho_0,\rho_\uptau)$ for five ratios of $\Delta_{K'}$ over $\Delta_K$, at $B=0$ (left side) and $B=5$ T (right side).
The purple (yellow) circles in (a), (b), (c) and (d) mark the smallest (largest) detuning difference $\Delta_d|_{\min(\max)}$.
We choose the detuning $\Delta_K=5$ meV.}
\label{figure5}
\end{figure}

We also show the dynamical evolution of population imbalance (i.e. $\delta\rho=\rho^{K}-\rho^{K'}$), for five ratios of $\Delta_{K'}$ over $\Delta_K$ at $B=0$ [Fig.~\ref{figure5}(a)] and $B=5$ T [Fig.~\ref{figure5}(b)].
For the symmetric excitation at zero magnetic field [Fig.~\ref{figure5}(a)], due to an initial coherent superposition of excitonic states with the same population in the $K$ and $K'$ valleys, the dynamical evolution of $\rho^K$ and $\rho^{K'}$ perfectly matches [see the black line in Fig.~\ref{figure5}(a)].
While for the asymmetric excitation, the detuning difference $\Delta_d$ leads to incongruous dynamical behaviors of $\rho^K$ and $\rho^{K'}$.
Regarding the case of $\Delta_K\textgreater\Delta_{K'}$, the population imbalance $\delta\rho$ increases from zero to the maximum when the time less than 1 ps.
After that, it drops over a long period of time until reaches the minimum [see the blue and red curves in Fig.~\ref{figure5}(a)].
The underlying physics can be understood as follows.
The coupling between the TMDC and laser field leads to a coherent transfer of excitonic population from the excitonic state $|\xi\rangle$ to the ground state $|0\rangle$.
Since we consider the two coupling strengths $g_K$ and $g_{K'}$ are the same, the larger detuning $\Delta_K$ disfavors the coherent transfer compare to the smaller one, which consequently results in $\rho^{K}$ decaying slower than $\rho^{K'}$ in the initial.
However, affected by the intervalley scattering and radiative recombination of excitons, the unequal distribution of excitonic population between the $K$ and $K'$ valleys is considerably quenched, giving a reduction to the population imbalance $\delta\rho$.
Considering the dynamical equilibrium of the system, an asymmetric excitation with $\Delta_K\textgreater\Delta_{K'}$ enables the coupling of the TMDC to laser field in $K'$ valley to be more efficient than that in $K$ valley, so that the $K'$ valley holds eventually more excitonic populations.
In contrast, when using an asymmetric excitation with $\Delta_K\textless\Delta_{K'}$, the population imbalance $\delta\rho$ has an opposite dynamical evolution [see the pink and green curves in Fig.~\ref{figure5}(a)].

In the presence of magnetic field, the dynamical evolution of population imbalance $\delta\rho$ under symmetric excitation varies from that at zero magnetic field [cf. the black lines in Figs.~\ref{figure5}(a) and~\ref{figure5}(b)].
The population imbalance $\delta\rho$ initially undergoes slight fluctuation around the zero and then grows to its maximum over time.
This variation arises from the combined effect of the detuning difference and asymmetric intervalley scattering on valley dynamics, both of which are induced by the valley splitting.
Specifically, since the $K$ valley is the lower one in energy, the asymmetric intervalley scattering pushes more excitonic populations to transfer to the $K$ valley.
However, in the initial stage of valley dynamics, the detuning difference $\Delta_d$ slows down the coherent transfer from the state $|K'\rangle$ to the state $|0\rangle$, which is responsible for the slight dropping of population imbalance $\delta\rho$ when the time less than 1 ps.
Also, under asymmetric excitation, the dynamical evolutions of population imbalance $\delta\rho$ shift upward with respect to those at zero magnetic field, since the intervalley scattering inhibits the incoherent transfer of excitons from $K$ valley to $K'$ valley in the presence of magnetically-induced valley splitting [cf. the green, pink, red and blue curves in Figs.~\ref{figure5}(a) and~\ref{figure5}(b)].

\subsection{Two optimal schemes: intrinsic distinction and practical application}

\begin{figure}
\includegraphics[width=\linewidth]{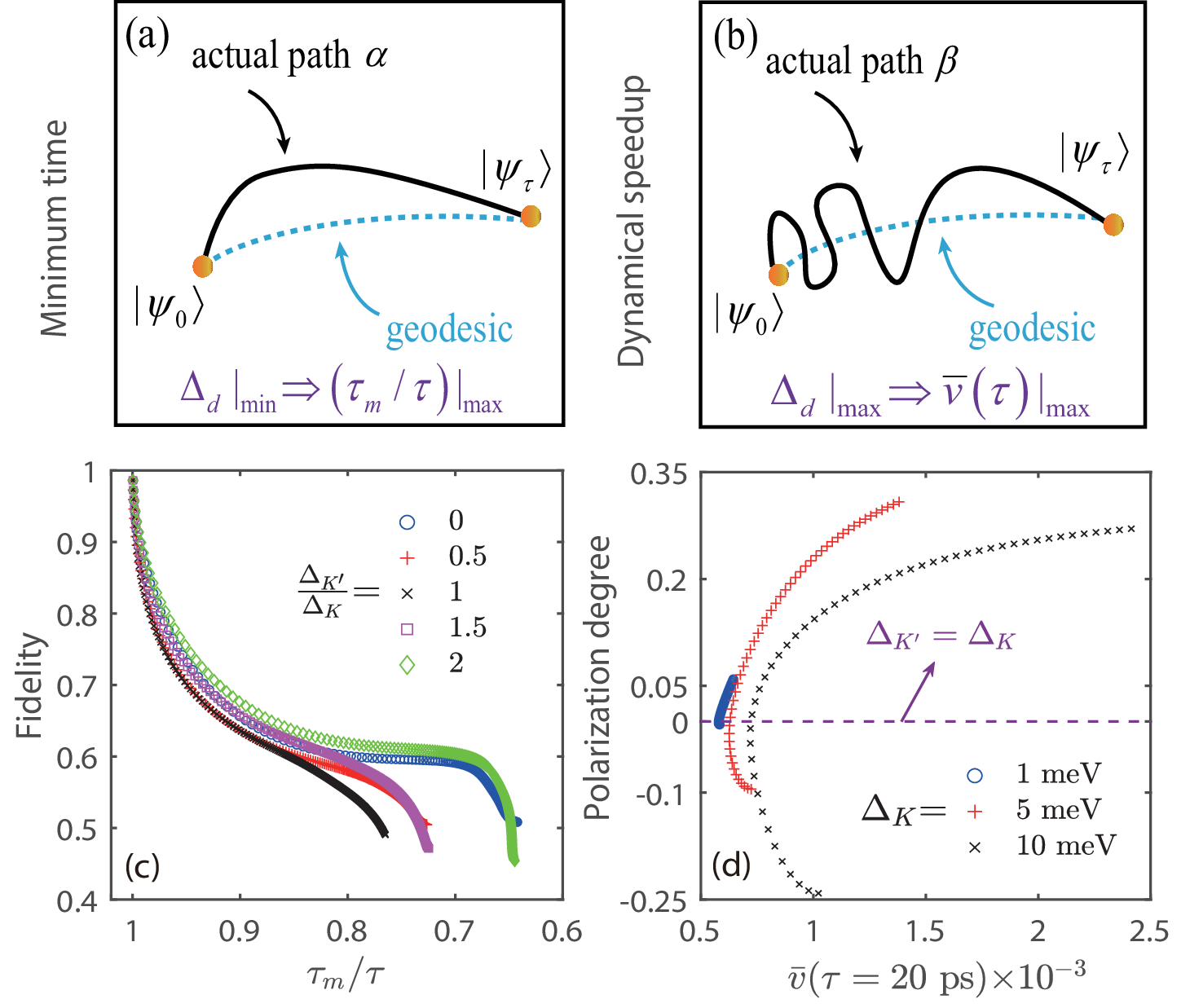}
\caption{Schematic of two valley dynamical pathways between an initial state $|\uppsi_0\rangle$ and a final state $|\uppsi_\uptau\rangle$ that devote to reduce the actual evolution time (a) and to boost the evolution speed (b), respectively.
The trajectory $\alpha$ denotes an optimal one which has the maximum value of ratio $\uptau_{\rm{m}}/\uptau$ for all potential evolution paths, while the trajectory $\beta$ indicates the fastest evolving one with the maximum average speed $\bar v(\uptau)$.
(c) The relationship between the system fidelity and the ratio $\uptau_{\rm{m}}/\uptau$ from $\uptau=0$ to $\uptau=10$ ps for five ratios of $\Delta_{K'}$ over $\Delta_K$.
(d) The relationship between the valley polarization and average evolution speed from $\Delta_{K'}=0$ to $\Delta_{K'}=4\Delta_K$ with different detuning $\Delta_K$.
The purple dashed line denotes the valley polarization $\eta=0$ when using a symmetric excitation with $\Delta_K=\Delta_{K'}$.
The upper (lower) side of the purple dashed line indicates that the valley polarization $\eta\textgreater0$ ($\eta\textless0$) when using an asymmetric excitation with $\Delta_K\textless\Delta_{K'}$ ($\Delta_K\textgreater\Delta_{K'}$).
In (c), we choose the detuning $\Delta_K=5$ meV.
In (d), we consider the actual evolution time $\uptau=20$ ps.}
\label{figure6}
\end{figure}

In this subsection, we further explore the intrinsic distinction between two types of control schemes of valley dynamics.
Benefiting from our dynamical analysis of excitonic population [Figs.~\ref{figure5}(a) and~\ref{figure5}(b)], we find that the optimal control devoted to the reduction of the evolution time [Fig.~\ref{figure2}(c)], requires that population imbalance $\delta\rho$ converges to zero [see the black line in Fig.~\ref{figure5}(a)], in which the detuning difference is the smallest [see the purple circle in Fig.~\ref{figure5}(a)].
Hence, it is essential to reduce the evolution time of valley dynamics by limiting the detuning difference.
In the absence of magnetic field, the symmetric excitation with $\Delta_K=\Delta_{K'}$ leads to a consequence of detuning difference $\Delta_d=0$.
While in the presence of magnetic field, the valley splitting effectively quenches the detuning difference arising from asymmetric excitation.
We further show the actual evolution length $\mathcal{L}(\rho_0,\rho_\uptau)$ of valley dynamics at $B=0$ [Fig.~\ref{figure5}(c)] and $B=5$ T [Fig.~\ref{figure5}(d)].
It can be found that the actual evolution length of the valley dynamics is the shortest where the detuning difference is the smallest [see the black curve in Fig.~\ref{figure5}(c) and red curve in Fig.~\ref{figure5}(d)].
To facilitate the understanding, we plot a schematic to illustrate this dynamical path [Fig.~\ref{figure6}(a)], in which the system evolves from an initial state $|\uppsi_0\rangle$ along a \emph{smooth} curve to reach the target state $|\uppsi_{\uptau}\rangle$, i.e., a path that converges towards the geodesic path.

Regrading the enhancement of average evolution speed, we find that a faster dynamical evolution [see the red circle in Fig.~\ref{figure4}(d)] leads to a significant population imbalance $\delta\rho$ [see the green curve in Fig.~\ref{figure5}(b)], in which, the detuning difference is the largest [see the yellow circle in Fig.~\ref{figure5}(b)].
In the absence of magnetic field, the asymmetric excitation with $\Delta_K\not=\Delta_{K'}$ induces an emerging of detuning difference.
In addition, the detuning difference can be further enhanced by the synergistic effect of the zero-field detuning and valley splitting.
Also, with the increase of detuning difference $\Delta_d$, the evolution path of valley dynamics is accordingly lengthened, thereby distancing it from the geodesic path [see the green curves in Figs.~\ref{figure5}(c) and~\ref{figure5}(d)].
This can be understood as follows: The perfectly equivalent initial distribution of excitonic population ($\delta\rho=0$) fails to match the energy level configuration ($\Delta_d\not=0$), which pushes the system to take a fast evolution to overcome this mismatch, accompanied by a pronounced excitonic population redistribution between the $K$ and $K'$ valleys.
For completeness, we show the schematic to illustrate the dynamical path with a greater evolution speed [Fig.~\ref{figure6}(b)], in which the system evolves along a \emph{complex} curve, i.e., a path that departs from the geodesic path.

Next, we demonstrate the practical application of two optimal control schemes of valley dynamics in the system fidelity $\mathcal{F}$ and valley polarization $\eta$.
In the transmission of quantum information, there are unavoidable errors, and the research on the problem of information correction has accordingly been arisen.
The investigation of the dynamical decay of system fidelity may help us understand the causes of errors related to various parameters in quantum operation, and contribute to the improved algorithms.
In general, the fidelity was originally introduced as a transition probability between two states, $\mathcal{F}_t=\rm{Tr}(\sqrt{\rho_0}\rho_t\sqrt{\rho_0})^{\frac{1}{2}}$~\cite{1994JMOp,PhysRevLett.110.050402}.
Since the linearly polarized excitation generates an initial pure state, the fidelity can be reduced as $\mathcal{F}_t=\sqrt{\langle\uppsi_0|\rho_t|\uppsi_0\rangle}=\sqrt{\rm{Tr}(\rho_0\rho_t)}$~\cite{PhysRevA.94.052125,PhysRevLett.111.010402}.
Figure~\ref{figure6}(c) shows the relationship between the system fidelity $\mathcal{F}$ and the ratio $\uptau_{\rm{m}}/\uptau$ from $\uptau=0$ to $\uptau=10$ ps.
For completeness, we consider five ratios of $\Delta_{K'}$ over $\Delta_K$.
Firstly, we find a sharp drop in the fidelity $\mathcal{F}$ as the ratio $\uptau_{\rm{m}}/\uptau$ decreases.
After that, the fidelity $\mathcal{F}$ holds stability and then recovers the drop.
High value of fidelity emerges accompanied by a large ratio, indicating that there is a weak information loss where the system evolves along a path taking less time.
Note that, a larger detuning difference allows the fidelity $\mathcal{F}$ to hold stability over a wider range [see the blue circle (\textcolor{blue}{$\circ$}) and green diamond (\textcolor{green}{$\diamond$}) markers in Fig.~\ref{figure6}(c)], which effectively suppress information loss in the process of actual evolution path deviating from the geodesic path.
The close association between system fidelity $\mathcal{F}$ and ratio $\uptau_{\rm{m}}/\uptau$ indicates that information loss in the valley dynamical evolution can be prevented by pushing the actual evolution time to converge to the QSL time.

The valley polarization is a useful feature for signal readout of valley information, manifested by the photoluminescence (PL) spectra with the opposite circularly polarized emissions ($\sigma_+$ and $\sigma_-$), i.e., $\eta=(I_{\sigma_+}-I_{\sigma_-})/(I_{\sigma_+}+I_{\sigma_-})$~\cite{PhysRevB.97.115425}, where $I_{\sigma_+}$ and $I_{\sigma_-}$ are the PL intensities of the $\sigma_+$ and $\sigma_-$ components, respectively.
General approaches for creating and manipulating valley polarization, such as external magnetic field and intense circularly polarized excitation~\cite{aivazianmagnetic2015,cao2012}, were developed to spread out the valleytronic application.
Figure~\ref{figure6}(d) shows the relationship between the valley polarization $\eta$ and evolution speed $\bar v(\uptau)$ from the $\Delta_{K'}=0$ to $\Delta_{K'}=4\Delta_K$.
Note that, there is no initial valley polarization with respect to the initial coherent superposition of excitoninc states we considered.
With fixing the detuning $\Delta_K$, we find that the valley polarization $\eta$ can be generated and further enhanced as the evolution speed $\bar v(\uptau)$ increases.
In addition, with the increase of detuning $\Delta_K$, though the evolution speed is effectively raised, it does not automatically signal that the valley polarization is strengthened consequently, cf. the red plus (\textcolor{red}{$+$}) and black cross ($\times$) markers in Fig.~\ref{figure6}(d).
Nevertheless, we emphasize that the fast evolving valley dynamics might yield some insights for enhancing valley polarization.

\section{Conclusions}
\label{sec:conc}

To endow an explicit optimal control scheme of valley dynamics, we construct a comprehensive model that incorporates both intra- and intervalley channels of excitons in monolayer WSe$_2$, and simultaneously take the light-matter interaction into account.
In light of the geometric QSL theory, we propose two optimal control schemes for optimally tuning valley dynamics, which are respectively devoted to reduce the evolution time of reaching the target state, and to boost the evolution speed over a period of time.
We reveal that the detuning difference has an essential influence on the implementation of dynamical optimization, allowing to control the evolution path of valley dynamics by means of the optical excitation mode and external magnetic field.
Also, we demonstrate that two optimal control schemes may offer insight into maintaining high fidelity in the information transmission, as well as boosting valley polarization, respectively.
The effect of magnetic field induced valley Zeeman splitting has been also discussed.

\setlength{\parskip}{-0.058cm}
As a remark, though the two types of schemes ask different requirements for the detuning difference, the physical sources underlying them are not contradictory.
This affords to tune valley dynamics in various perspective for diversified targets, since the resulting control closely depends on the \emph{performance measure} to be optimized~\cite{Deffner_2014}.
For practical considerations, our results show that the optical excitation with resonance detuning (i.e. $\Delta_\xi=0$) is not necessary for reducing the actual evolution time of valley dynamics.
This relaxes the requirement for the coupling performance between the TMDC and optical cavity.
Additionally, the nonuniqueness of a bona fide metric defining the geodesic reminds to reinforce further exploration of improved QSL bounds~\cite{PhysRevX.6.021031}.
Our research generalizes the practical application of QSL theory, and also provides an effective strategy for optically regulating the dynamical evolution in valley qubit.

\section*{Acknowledgments}
We thank X.-J. Cai for valuable discussions.
This work was supported by the National Natural Science Foundation of China (Nos. 11974212, 12274256, and 11874236) and the Major Basic Program of
Natural Science Foundation of Shandong Province (Grant No. ZR2021ZD01).
\end{CJK}

\begin{thebibliography}{90}
\bibitem [{foo()}]{foot1}%
  \BibitemOpen
  \href@noop {} {}\bibinfo {note} {Since we consider a weak coupling between
  the TMDC with optical field, the dissipation could still be regarded as along
  the direction of $|\xi\rangle$ to $|0\rangle$, see the similar treatments in
  Ref.
  [\onlinecite{PhysRevB.81.075322,PhysRevB.107.035404,PhysRevA.68.012310}].}\BibitemShut
  {Stop}%
\end{thebibliography}
%

\end{document}